\documentstyle[amstex,amssymb,aps]{revtex}

\begin{document}
\author{M. Sirera and A. P\'{e}rez.}
\address{Departamento de F\'{\i }sica Te\'{o}rica, Universidad de Valencia \\
E-46100 Burjassot (Valencia) Spain.}
\title{Relativistic Wigner Function Approach to Neutrino Propagation in Matter.}
\maketitle

\begin{abstract}
In this work we study the propagation of massive Dirac neutrinos in matter
with flavor mixing, using statistical techniques based on Relativistic
Wigner Functions. First, we consider neutrinos in equilibrium within the
Hartree approximation, and obtain the corresponding relativistic dispersion
relations and effective masses. After this, we analyze the same system out
of equilibrium. We verify that, under the appropiate physical conditions,
the well known equations for the MSW effect are recovered. The techniques we
used here appear as an alternative to describe neutrino properties and
transport equations in a consistent way.
\end{abstract}

\section{Introduction}

Neutrino propagation in dense media becomes an important issue in some
astrophysical scenarios, such as supernovae, neutron stars during the
Kelvin-Helmholtz epoch and the solar neutrino problem \cite{Ba89}, \cite{R96}%
. The first two cases correspond to compact stars, where densities a few
times the nuclear saturation density are reached. To describe neutrino
propagation in the dense core of such compact objects, aside of production
and absorption rates\footnote{%
In this work we will concentrate on medium effects in neutrino propagation,
and will not discuss production and absortion mechanisms. For these
processes, the reader may want to consult \cite{Bru85}.}, one is mostly
interested in the neutrino cross section with the surrounding matter, which
in turn is related to the imaginary part of the forward amplitude, due to
the optical theorem. In the case of solar neutrinos (and also for the
'atmosphere' of compact stars), the possibility of neutrino oscillations and
flavor conversion appears. In the standard picture, these oscillations are
associated to the real part of the forward scattering amplitude of massive
neutrinos with the background.

In this paper we will be concerned about the latter topic, i. e. matter
effects on neutrino masses and neutrino oscillations. Since the seminal
papers by Wolfenstein \cite{Wo78},\cite{Wo79} and by Mikheyev and Smirnov 
\cite{MS86a},\cite{MS86b},\cite{MS86c}, there have been many papers which
have explored the physics of in-medium neutrinos, specially in connection
with the solar neutrino problem (for a review, see for example \cite{KP89}
and references \cite{Ba89}, \cite{R96}). Different techniques have been used
to approach this problem. The simplest one consists in describing the matter
effect by an {\it effective potential} (as in the Wolfenstein's papers)
which will be added to the mass matrix to give an {\it effective Hamiltonian}
in the Schr\"{o}dinger's equation. Although this approach will suffice for
most purposes, it is clear that it does not describe matter effects in a
covariant way. In order to obtain covariant equations, one has to obtain the
corresponding dispersion relations, which are the in-medium analogous to the
simple mass-shell condition $p^{2}=m^{2}$ of free particles. Dispersion
relations of neutrinos in different backgrounds, taking into account the
mixing among generations, have been investigated by N\"{o}tzold and Raffelt 
\cite{NR88}. In that work, dispersion relations appear as the poles of the
neutrino propagator (Green function), evaluated at the one-loop
approximation.

On the other hand, one has astrophysical situations in which contribution of
neutrinos to macroscopic magnitudes, such as the energy, pressure, etc, ...
becomes important (this is the case in a supernova collapse \cite{Bru85}, or
in the Early Universe). In this case, one has to introduce a {\it %
distribution function} for neutrinos to describe \ the number of neutrinos
having a given momentum. It is then desirable to develop evolution equations
for these functions in the case where neutrino oscillations are present,
which implies that distribution functions become non-diagonal matrices in
flavor space. Several works have implemented this in different ways. In Ref. 
\cite{Ru90}, use is made of the techniques described in \cite{Aj77} to
obtain the kinetic equations for non-relativistic Wigner distribution
functions of neutrinos. Alternatively, in references \cite{RSiSt93},\cite
{RS93} it is derived the time evolution of a neutrino density matrix $\rho $
constructed as macroscopic averages of generalized occupation numbers, in
the way $\rho _{ij}(\vec{p})=<\hat{\rho}_{ij}(\vec{p})>$ , where $\hat{\rho}%
_{ij}(\vec{p})=a_{j}^{+}(\vec{p})a_{i}(\vec{p})$ and $a_{j}^{+}(\vec{p})$ ($%
a_{i}(\vec{p})$) is the creation (destruction) operator of neutrinos with
flavor $j$ ($i$) and momentum $\vec{p}$. Here, one starts directly from
Heisenberg's equation 
\begin{equation}
i\partial _{t}\hat{\rho}=[\hat{\rho},H]  \label{Heisenberg}
\end{equation}
with a Hamiltonian $H=H_{0}+H_{int}$, where $H_{0}$ is the free Hamiltonian
and $H_{int}$ is the interaction piece. Equation (\ref{Heisenberg}) is then
expanded perturbatively (after macroscopic averaging). With this at hand,
the authors have studied the possibility of flavor conversion in a supernova
core (see the references above for more details).

Both procedures give rise to an expansion in powers of the Fermi's coupling
constant $G_{F}$. The first term in this expansion (proportional to $G_{F}$)
contains the modifications to the mass matrix due to the interaction, while
the second, $G_{F}^{2}$ term, is the generalization of the Boltzmann
collision integral to the case of flavor mixing. However, because both
methods are based on non-covariant techniques, one does not obtain
relativistic dispersion relations for the neutrinos.

In this paper, we make use of {\em relativistic} Wigner functions \cite{Ha78}%
,\cite{GLW80} to describe propagation of neutrinos with flavor mixing in
dense media. Relativistic Wigner functions have been successfully used to
account for finite density and temperature effects in nuclear matter \cite
{DH84}, \cite{Di84},\cite{Di85}. They provide an alternative to Green
functions in a way which is well adopted to the development of kinetic
equations. In addition to this, temperature effects are incorporated in a
unique way, contrarily to the situation of Green functions. We also need to
introduce correlation functions among neutrinos and the background (which,
for simplicity, is considered to consist only on electrons). These
correlations, as will be discussed later, take into account for the residual
interaction of neutrinos and electrons beyond the mean-field approximation.
We will obtain, in the next section, the equations of motion for these
functions, under the assumption that the interacting background is in an
equilibrium state.

\bigskip Next, we will consider some particular situations. In section 3, we
examine the case when neutrinos are in equilibrium and correlations are
neglected. The dispersion relations reproduce, in this case, the expected
effective masses and in-matter mixing angles. This result supports our
statement that neglecting correlations is equivalent to a mean-field
treatment of the surrounding matter.

In section 4 we examine some departure from the above simplest situation, by
keeping spatial and/or time variations in the kinetic equations (with
correlations still neglected), and we consider propagation on a
density-varying medium. In section 5 we examine with more detail a
particular case, corresponding to a small effective potential and
macroscopic inhomogeneities which are large enough. This is the situation
encountered when dealing with solar neutrinos. By making the appropriate
approximations, we recover the well-known formulae for the MSW effect. This
shows the ability of Wigner function techniques to correctly reproduce both
relativistic dispersion equations and transport equations on the same foot,
at least in the cases we have studied. We end in section 6 by summarizing
our main results and making some remarks. Some auxiliary results will be
given in the appendixes. The construction of the Wigner function in the case
of free neutrinos is showed in Appendix A. In Appendix B we analyze the
neutrino dispersion relations and effective masses which appear within the
Hartree approximation, and we extend the results of Appendix A to the
construction of the corresponding Wigner function.

In this work the metric is$\ g^{\mu \nu }=diag(1,-1,-1,-1)$. We adopt the
chiral representation for gamma matrices, and natural units ($\hbar =c=1$)
are used.

\section{Equations for Wigner functions}

In this section we derive the equations of motion for neutrino Wigner
functions. Neutrinos are assumed to propagate on a matter background in
equilibrium, an hypothesis which holds in all the astrophysical scenarios
mentioned in the introduction.

In order so simplify the equations as much as possible, we consider only two
neutrino flavors (namely electron and muon neutrinos) . Our treatment can be
generalized in a straightforward way to include more neutrino flavors. Also,
our main concern in neutrino propagation is neutrino oscillations and flavor
conversion, which are given by charged-current interactions with electrons.
Therefore, we consider a background consisting on electrons, although we
take into account both neutral and charged currents. As before, the
formalism can be trivially extended to account for a more general situation,
such as neutral-current interactions on protons and neutrons. In this paper,
we treat neutrinos as massive Dirac particles in the most simple model for
massive neutrinos, i.e. we treat them the same way as the all other fermions
(leptons and quarks). Within this minimal extension of The Standard
Electroweak Theory, the charge of the neutrino field which is conserved is
the total lepton number $L=L_{e}+L_{\mu }$. On the other hand, as we deal
with low-energy neutrinos (with energies of the order a few MeV), we adopt
an effective contact interaction{\em \ } of neutrinos with the matter
background.

In what follows, neutrino magnitudes without a prime will indicate flavor
states, and primes will be used for {\em free mass eigenstates}. In the next
sections and in the appendixes, a tilde will be used for {\em interacting
eigenstates. }Flavors, as well as mass eigenstates, will be labeled by
lattin superindexes such as $a$ and $b$. Spin subindexes will be generally
omitted ; when needed, we use indices such as $i,j,k,...$ to label them.
Lorentz indices will be labeled by Greek letters.

Since we deal with two neutrino species, it is convenient to introduce
vectors and matrices in flavor (or mass) space. We therefore define the
neutrino and antineutrino vector fields\footnote{%
The symbol $\ \symbol{94}$ on top of a magnitude means that we are dealing
with a quantum operator. This will be used to distinguish this magnitudes
from statisical averages.} : 
\begin{equation}
\widehat{\nu }(x)\equiv \left( 
\begin{array}{c}
\widehat{\nu }^{e}(x) \\ 
\widehat{\nu }^{\mu }(x)
\end{array}
\right) ,\ \ \widehat{\overline{\nu }}(x)\equiv \left( 
\begin{array}{cc}
\widehat{\overline{\nu }}^{e}(x) & \widehat{\overline{\nu }}^{\mu }(x)
\end{array}
\right)  \label{21}
\end{equation}

We also introduce the following matrices in flavor space 
\begin{eqnarray}
\Lambda ^{ab,\mu } &\equiv &\left( 
\begin{array}{cc}
0 & 0 \\ 
0 & \lambda ^{\mu }
\end{array}
\right)  \label{24} \\
\quad \Omega ^{ab,\mu } &\equiv &\left( 
\begin{array}{cc}
\omega ^{\mu } & 0 \\ 
0 & 0
\end{array}
\right)
\end{eqnarray}
With the notations $\lambda ^{\mu }\equiv \gamma ^{\mu }(g_{V}-g_{A}\gamma
^{5})$ and $\omega ^{\mu }\equiv \gamma ^{\mu }(\widetilde{g}_{V}-\widetilde{%
g}_{A}\gamma ^{5})$. Here, the constant $g_{V}=-\frac{1}{2}+2\sin ^{2}\theta
_{W}$ ($g_{A}=-\frac{1}{2}$ ) corresponds to the vector (axial) contribution
of weak neutral currents, $\theta _{W}$ is the Weinberg's angle, while the
constants $\widetilde{g}_{V}=\frac{1}{2}+2\sin ^{2}\theta _{W}$ and $%
\widetilde{g}_{A}=\frac{1}{2}$ arise from neutral plus charged currents.

With these notations, the Lagrangian density is written as :

\begin{equation}
\widehat{{\cal L}}(x)=\widehat{{\cal L}}_{e}(x)+\widehat{{\cal L}}_{\nu }(x)+%
\widehat{{\cal L}}_{I}(x)  \nonumber
\end{equation}

with $\widehat{{\cal L}}_{e}(x)$ the Lagrangian of free electrons, 
\begin{equation}
\widehat{{\cal L}}_{\nu }(x)=\widehat{\overline{\nu }}(x)i\gamma ^{\mu
}\partial _{\mu }\widehat{\nu }(x)-\widehat{\overline{\nu }}(x)M\widehat{\nu 
}(x)  \label{lfreenu}
\end{equation}

the corresponding Lagrangian of free neutrinos, and 
\begin{equation}
\widehat{{\cal L}}_{I}(x)=-\frac{G_{F}}{\sqrt{2}}\widehat{\overline{\nu }}%
(x)\Omega ^{\mu }\widehat{\nu }(x)\widehat{\bar{e}}(x)\omega _{\mu }\widehat{%
e}(x)-\frac{G_{F}}{\sqrt{2}}\widehat{\overline{\nu }}(x)\Lambda ^{\mu }%
\widehat{\nu }(x)\widehat{\bar{e}}(x)\lambda _{\mu }\widehat{e}(x)
\end{equation}

the interaction piece. In Eq. (\ref{lfreenu}) $M$ is the free mass matrix 
{\em in flavor space} \footnote{%
If flavor mixing exists, $M$ is non-diagonal, its eigenvalues being the
masses of free mass eigenstates.}. From the above Lagrangian one readily
obtains the equations of motion for the neutrinos : 
\begin{equation}
i\gamma ^{\mu }\partial _{\mu }\widehat{\nu }(x)-M\widehat{\nu }(x)-\frac{%
G_{F}}{\sqrt{2}}\Omega ^{\mu }\widehat{\nu }(x)\widehat{\overline{e}}%
(x)\omega _{\mu }\widehat{e}(x)-\frac{G_{F}}{\sqrt{2}}\Lambda ^{\mu }%
\widehat{\nu }(x)\widehat{\overline{e}}(x)\lambda _{\mu }\widehat{e}(x)=0
\label{fieldeqa}
\end{equation}
\begin{equation}
\partial _{\mu }\widehat{\overline{\nu }}(x)i\gamma ^{\mu }+\widehat{%
\overline{\nu }}(x)M+\frac{G_{F}}{\sqrt{2}}\widehat{\overline{\nu }}%
(x)\Omega ^{\mu }\widehat{\overline{e}}(x)\omega _{\mu }\widehat{e}(x)+\frac{%
G_{F}}{\sqrt{2}}\widehat{\overline{\nu }}(x)\Lambda ^{\mu }\widehat{%
\overline{e}}(x)\lambda _{\mu }\widehat{e}(x)=0  \label{fieldeqb}
\end{equation}

We now introduce the neutrino Wigner {\em operator} 
\begin{equation}
\widehat{F}_{ij}^{ab(\nu )}(x,p)=(2\pi )^{-4}\int d^{4}y\;e^{-ipy}\,\widehat{%
\overline{\nu }}_{j}^{b}(x+y/2)\widehat{\nu }_{i}^{a}(x-y/2)
\label{wigopneu}
\end{equation}

and the electron Wigner operator 
\begin{equation}
\widehat{F}_{ij}^{(e)}(x,p)\equiv (2\pi )^{-4}\int d^{4}y\;e^{-ipy}\widehat{%
\,\overline{e}}_{j}(x+y/2)\widehat{e}_{i}(x-y/2)
\end{equation}
From Eq. (\ref{wigopneu}) one readily gets that the Hermitian conjugate is
given by 
\begin{equation}
\widehat{F}_{ij}^{ab(\nu )\dagger }(x,p)=\gamma _{jk}^{0}\widehat{F}%
_{kq}^{ba(\nu )}(x,p)\gamma _{qi}^{0}  \label{216}
\end{equation}

Here and hereafter summation over repeated indices is understood. With the
help of Eqs. (\ref{fieldeqa}) and (\ref{fieldeqb}) one can obtain the
equations of motion obeyed by the neutrino Wigner operators. After some
algebra, we get :

\begin{eqnarray}
&&\gamma \left[ \partial \widehat{F}^{(\nu )}(x,p)-2ip\widehat{F}^{(\nu
)}(x,p)\right] +2iM\widehat{F}^{(\nu )}(x,p)=  \label{eqwiga} \\
&&\frac{-2iG_{F}}{\sqrt{2}}(2\pi )^{-4}\int d^{4}yd^{4}k\left[ \Omega 
\widehat{\bar{e}}(y)\omega \widehat{e}(y)+\Lambda \widehat{\bar{e}}%
(y)\lambda \widehat{e}(y)\right] \widehat{F}^{(\nu )}(x,p-k/2)e^{ik(y-x)} 
\nonumber
\end{eqnarray}
\begin{eqnarray}
&&\left[ \partial \widehat{F}^{(\nu )}(x,p)+2ip\widehat{F}^{(\nu )}(x,p)%
\right] \gamma -2iM\widehat{F}^{(\nu )}(x,p)=  \label{eqwigb} \\
&&\frac{2iG_{F}}{\sqrt{2}}(2\pi )^{-4}\int d^{4}yd^{4}k\widehat{F}^{(\nu
)}(x,p-k/2)\left[ \Omega \widehat{\bar{e}}(y)\omega \widehat{e}(y)+\Lambda 
\widehat{\bar{e}}(y)\lambda \widehat{e}(y)\right] e^{-ik(y-x)}  \nonumber
\end{eqnarray}

One could also derive the corresponding equations for the electron Wigner
operator $\widehat{F}_{ij}^{(e)}(x,p)$. However, we will not need these
equations under the approximations discussed in this paper, and therefore we
will omit them. Of course, if one wants to investigate the next order to
this approximation, the whole system of equations has to be taken into
account.

We are now interested in introducing statistical averages from the quantum
operators introduced above. These statistical averages are called Wigner
functions \cite{W32}, and are the analogous to the distribution functions we
need to describe many-particles systems. These are, in general, complex
functions, and also contain a Lorentz structure which will be discussed
later. The electron and neutrino Wigner functions are defined, respectively,
as : 
\begin{equation}
F_{ij}^{(e)}(x,p)\equiv <\widehat{F}_{ij}^{(e)}(x,p)>=(2\pi )^{-4}\int
d^{4}y\,e^{-ipy}<\widehat{\bar{e}}_{j}(x+y/2)\widehat{e}_{i}(x-y/2)>
\label{wigfel}
\end{equation}
\begin{equation}
F_{ij}^{ab(\nu )}(x,p)\equiv <\widehat{F}_{ij}^{ab(\nu )}(x,p)>=(2\pi
)^{-4}\int d^{4}y\,e^{-ipy}<\widehat{\bar{\nu}}_{j}^{b}(x+y/2)\widehat{\bar{%
\nu}}_{i}^{a}(x-y/2)>  \label{wigfne}
\end{equation}

Here, the symbol $<\widehat{A}>$ means the average of a given quantum
operator $\widehat{A}$ over a basis of quantum states which are compatible
with the macroscopical knowledge of the system. The latter determines a
given density matrix operator $\widehat{\rho }$ . Thus the averaging is
performed according to 
\begin{equation}
<\widehat{A}>\equiv Sp\{\widehat{\rho }\widehat{A}\}  \label{average}
\end{equation}

where $Sp$ means the trace performed over the quantum basis. By taking this
average on Eqs. (\ref{eqwiga}) and (\ref{eqwigb}) one arrives to the
following equations : 
\begin{eqnarray}
&&\left[ \gamma (\partial -2ip)+2iM\right] F^{(\nu )}(x,p)=-\frac{iG_{F}%
\sqrt{2}}{(2\pi )^{4}}\int d^{4}yd^{4}kd^{4}k^{\prime }e^{ik(y-x)}  \nonumber
\\
\lbrack <\Omega Tr(\omega \widehat{F}^{(e)}(y,k^{\prime })) &&\widehat{F}%
^{(\nu )}(x,p-k/2)>+<\Lambda Tr(\lambda \widehat{F}^{(e)}(y,k^{\prime }))%
\widehat{F}^{(\nu )}(x,p-k/2)>]  \label{wfneua} \\
&&F^{(\nu )}(x,p)\left[ \gamma (\partial +2ip)-2iM\right] =\frac{iG_{F}\sqrt{%
2}}{(2\pi )^{4}}\int d^{4}yd^{4}kd^{4}k^{\prime }e^{-ik(y-x)}  \nonumber \\
\lbrack <\widehat{F}^{(\nu )}(x,p-k/2) &&\Omega Tr(\omega \widehat{F}%
^{(e)}(y,k^{\prime }))>+<\widehat{F}^{(\nu )}(x,p-k/2)\Lambda Tr(\lambda 
\widehat{F}^{(e)}(y,k^{\prime }))>]  \label{wfneub}
\end{eqnarray}

In the latter equations, the symbol $Tr$ means the trace in spin indices.

Let us now introduce the electron-neutrino $A^{(\nu e)}$and
neutrino-electron $B^{(e\nu )}$correlation functions : 
\begin{eqnarray}
A_{ijkl}^{(\nu e)ab}(x,x^{\prime },p,p^{\prime }) &\equiv &<\widehat{F}%
_{ij}^{(\nu )ab}(x,p)\widehat{F}_{kl}^{(e)}(x^{\prime },p^{\prime
})>-F_{ij}^{(\nu )ab}(x,p)F_{kl}^{(e)}(x^{\prime },p^{\prime })
\label{corrnue} \\
B_{ijkl}^{(e\nu )ab}(x,x^{\prime },p,p^{\prime }) &\equiv &<\widehat{F}%
_{ij}^{(e)}(x,p)\widehat{F}_{kl}^{(\nu )ab}(x^{\prime },p^{\prime
})>-F_{ij}^{(e)}(x,p)F_{kl}^{(\nu )ab}(x^{\prime },p^{\prime })
\label{correnu}
\end{eqnarray}

Then Eqs. (\ref{wfneua}) and (\ref{wfneub}) can be rewritten as 
\begin{eqnarray}
&&\left[ \gamma (\partial -2ip)+2iM\right] F^{(\nu )}(x,p)=-\frac{iG_{F}%
\sqrt{2}}{(2\pi )^{4}}\int d^{4}yd^{4}kd^{4}k^{\prime }e^{ik(y-x)}  \nonumber
\\
\lbrack \Lambda Tr(\lambda &&B(y,x,k^{\prime },p-k/2))+\Omega Tr(\omega
B(y,x,k^{\prime },p-k/2))+  \nonumber \\
&&\Lambda Tr(\lambda F^{(e)}(y,k^{\prime }))F^{(\nu )}(x,p-k/2)+\Omega
Tr(\omega F^{(e)}(y,k^{\prime }))F^{(\nu )}(x,p-k/2)]  \label{wigcorra}
\end{eqnarray}
\begin{eqnarray}
&&F^{(\nu )}(x,p)\left[ \gamma (\partial +2ip)-2iM\right] =\frac{iG_{F}\sqrt{%
2}}{(2\pi )^{4}}\int d^{4}yd^{4}kd^{4}k^{\prime }e^{-ik(y-x)}  \nonumber \\
&&[\Lambda Tr(A(x,y,p-k/2,k^{\prime })\lambda )+\Omega
Tr(A(x,y,p-k/2,k^{\prime })\omega )+  \nonumber \\
&\Lambda &F^{(\nu )}(x,p-k/2)Tr(\lambda F^{(e)}(y,k^{\prime }))+\Omega
F^{(\nu )}(x,p-k/2)Tr(\omega F^{(e)}(y,k^{\prime }))]  \label{wigcorrb}
\end{eqnarray}

The two-point correlation functions defined above are not independent. In
fact, one can prove that they are related by : 
\begin{equation}
A_{ijkl}^{(\nu e)ab\ast }(x,x^{\prime },p,p^{\prime })=\gamma
_{lp}^{0}\gamma _{jr}^{0}B_{pqrs}^{(e\nu )ba}(x^{\prime },x,p^{\prime
},p)\gamma _{qk}^{0}\gamma _{si}^{0}
\end{equation}

which can be written, in a short way, as\footnote{%
For a matrix having both spin indices and generation indices, its Hermitian
conjugate is obtained by interchanging the generation indices and then
taking the Hermitian conjugate in spin space.} 
\begin{equation}
A^{(\nu e)\dagger }(x,x^{\prime },p,p^{\prime })=B^{(e\nu )}(x^{\prime
},x,p^{\prime },p)
\end{equation}

This implies that Eq. (\ref{wigcorrb}) is actually the Hermitian conjugate
of Eq. (\ref{wigcorra}).

\section{System in equilibrium. Hartree approximation}

In order to obtain some insight into the physical meaning of the neutrino
Wigner function we will, in this section, investigate the situation when
both the electrons and the neutrinos are in equilibrium, which we
characterize by all statistical magnitudes as being time-space
translationally invariant. This means that one-point functions can not
depend on $x$ , while two-point correlation functions can only depend on the
difference of coordinates, i. e. we assume that : 
\begin{eqnarray}
F^{(e)}(x,p) &=&F^{(e)}(p)\qquad  \label{cond1} \\
F^{(\nu )}(x,p) &=&F^{(\nu )}(p)
\end{eqnarray}

and 
\begin{eqnarray}
A^{(\nu e)}(x,x^{\prime },p,p^{\prime }) &=&A^{(\nu e)}(x-x^{\prime
},p,p^{\prime })  \nonumber \\
\ B^{(e\nu )}(x,x^{\prime },p,p^{\prime }) &=&B^{(e\nu )}(x-x^{\prime
},p,p^{\prime })  \label{cond2}
\end{eqnarray}

By defining the Fourier transform 
\begin{equation}
\widetilde{A}^{(\nu e)}(k,p,p^{\prime })=(2\pi )^{-4}\int
d^{4}x\,e^{-ikx}A^{(\nu e)}(x,p,p^{\prime })  \label{34}
\end{equation}

(analogously for $B^{(e\nu )}$), we obtain the equilibrium equations for the
neutrino Wigner function 
\begin{eqnarray}
&&\left[ p\gamma -M-\frac{G_{F}}{\sqrt{2}}\int d^{4}k^{\prime }\left(
\Lambda Tr\left( \lambda F^{(e)}(k^{\prime })\right) +\Omega Tr\left( \omega
F^{(e)}(k^{\prime })\right) \right) \right] F^{(\nu )}(p)=  \nonumber \\
&&\frac{G_{F}}{\sqrt{2}}\int d^{4}kd^{4}k^{\prime }\left[ \Lambda Tr\left(
\lambda \widetilde{B}^{(e\nu )}(k,k^{\prime },p+k/2)\right) +\Omega Tr\left(
\omega \widetilde{B}^{(e\nu )}(k,k^{\prime },p+k/2)\right) \right]
\label{eqwigeqa}
\end{eqnarray}
\begin{eqnarray}
&&F^{(\nu )}(p)\left[ p\gamma -M-\frac{G_{F}}{\sqrt{2}}\int d^{4}k^{\prime
}\left( Tr\left( \lambda F^{(e)}(k^{\prime })\right) \Lambda +Tr\left(
\omega F^{(e)}(k^{\prime })\right) \Omega )\right) \right] =  \nonumber \\
&&\frac{G_{F}}{\sqrt{2}}\int d^{4}kd^{4}k^{\prime }\left[ \Lambda Tr\left( 
\widetilde{A}^{(\nu e)}(k,p+k/2,k^{\prime })\lambda \right) +\Omega Tr\left( 
\widetilde{A}^{(\nu e)}(k,p+k/2,k^{\prime })\omega \right) \right]
\label{eqwigeqb}
\end{eqnarray}

As before, Eq. (\ref{eqwigeqb}) turns out to be the Hermitian conjugate of
Eq. (\ref{eqwigeqa}). This set of equations is obviously not complete, and
one should add the equations which are satisfied by the correlation
functions in looking for such a complete set. However, in doing so there
automatically appear {\em three-point} correlation functions. This procedure
can be infinitely continued, so that one obtains, instead of a closed set,
an infinite {\em hierarchy} similar to the BBGKY (after Bogoliubov-Born-
Green- Kirkwood-Yvon) hierarchy of classical systems \cite{Bo62},\cite{R80}.
For classical systems, one usually truncates this infinite chain by
neglecting correlations of order higher than a given one, usually by showing
that higher orders correspond to more rapid variations in space and time.
The next step consists then in incorporating perturabatively the next-order
correlations. We will use here the analogy with the classical situation, and
will first examine the situation at the lowest order, i. e. when all kind of
correlations are neglected. Such approximation is commonly referred to as
the {\em Hartree approximation. }As we will show, the neutrino dispersion
relations which arise from this approximation correspond to modifications of
the neutrino propagator at the one-loop level \cite{NR88}.

The Wigner function of electrons in equilibrium can be calculated using
standard techniques. Following \cite{HaHe78} one has 
\begin{equation}
F^{(e)}(p)=(2\pi )^{-3}\delta (p^{2}-m_{e}^{2})\left[ \theta
(p^{0})f_{e}^{+}(p)+\theta (-p^{0})f_{e}^{-}(p)\right] (\gamma p+m_{e})
\label{wigeqel}
\end{equation}

with $m_{e}$ the electron mass, $p^{0}$ the time-like component of the
electron four-momentum $p$ and $\theta (x)$ the step function. The functions 
$f_{e}^{+}(p)$, $f_{e}^{-}(p)$ are the Fermi-Dirac occupation numbers of
electrons and positrons, respectively. In the frame where the matter fluid
is at rest, they read as\footnote{%
Making the hypothesis that matter is at rest is equivalent to consider a
particular Lorentz frame such that the fluid four-velocity is $u^{\mu
}=(1,0,0,0)$. Results in other frames can be obtained by restoring the
four-velocity $u^{\mu }$, as discussed in \cite{GLW80}.} : 
\begin{equation}
f_{e}^{\pm }(p)=\frac{1}{e^{\beta (E\mp \mu _{e})}+1}
\end{equation}

(in Appendix A a similar calculation is shown for neutrinos) , where $E=%
\sqrt{\vec{p}^{2}+m_{e}^{2}}$, $\mu _{e}$ is the electron chemical potential
and $\beta $ the inverse temperature (we set the Boltzmann constant $k_{B}=1$%
). We can now calculate the traces and integrals appearing in Eqs. (\ref
{eqwigeqa}),(\ref{eqwigeqb}). After some algebra, it is easily obtained 
\begin{equation}
Tr\int d^{4}k\lambda ^{\mu }F^{(e)}(k)=\left\{ 
\begin{array}{c}
Tr\int d^{4}k\lambda ^{0}F^{(e)}(k)=g_{v}n \\ 
Tr\int d^{4}k\lambda ^{i}F^{(e)}(k)=0;\quad i=1,2,3
\end{array}
\right.  \label{382}
\end{equation}
Analogously 
\begin{equation}
Tr\int d^{4}k\omega ^{\mu }F^{(e)}(k)=\left\{ 
\begin{array}{c}
Tr\int d^{4}k\omega ^{0}F^{(e)}(k)=\widetilde{g}_{v}n \\ 
Tr\int d^{4}k\omega ^{i}F^{(e)}(k)=0;\quad i=1,2,3
\end{array}
\right.  \label{383}
\end{equation}
Let us now define the matrix (in flavor space) : 
\begin{equation}
\Phi =\left( 
\begin{array}{cc}
\widetilde{V} & 0 \\ 
0 & V_{n}
\end{array}
\right) .  \label{phimatrix}
\end{equation}

here, $V_{n}=\sqrt{2}G_{F}g_{v}n$ is the effective potential for neutral
currents, and$\widetilde{\text{ }V}=V+V_{n}$ , where $V=\sqrt{2}G_{F}n$ is
the corresponding potential for charged currents, with $n=4\int d^{4}k(2\pi
)^{-3}\delta (k^{2}-m_{e}^{2})(\theta (p^{0})f_{e}^{+}(k)+\theta
(-p^{0})f_{e}^{-}(k))k^{0}\equiv n_{e}-n_{\bar{e}\text{ }}$ the electron
(minus positron) number density. With these notations, and neglecting
correlations, Eq. (\ref{eqwigeqa}) can be cast under the form : 
\begin{equation}
\lbrack \gamma p-M-\gamma ^{0}\frac{1}{2}(1-\gamma ^{5})\Phi ]F(p)=0
\label{wigeqneu}
\end{equation}

(since electrons have been integrated out, the neutrino superscript in
Wigner functions will be omitted in what follows, in order to make notations
simpler). It is easily recognized in Eq. (\ref{wigeqneu}) the appearance of
the left-handed chirality projector $P_{L}=\frac{1}{2}(1-\gamma ^{5})$, as a
consequence of left-handed interactions. Let us also introduce the
right-handed projector $P_{R}=\frac{1}{2}(1+\gamma ^{5})$. With the help of
these two projectors, one can define the following components of the
neutrino Wigner function : 
\begin{eqnarray}
&&F_{L}=P_{L}FP_{R}  \nonumber \\
&&F_{R}=P_{R}FP_{L}  \nonumber \\
&&F_{RL}=P_{R}FP_{R}  \nonumber \\
&&F_{LR}=P_{L}FP_{L}  \label{projection}
\end{eqnarray}

If we apply $P_{L}$ and $P_{R}$ on the left and right of Eq. (\ref{wigeqneu}%
) we arrive to the set of equations 
\begin{eqnarray}
\gamma pF_{RL}-MF_{L} &=&0  \nonumber \\
\gamma pF_{LR}-MF_{R}-\gamma ^{0}\Phi F_{LR} &=&0  \nonumber \\
\gamma pF_{R}-MF_{LR} &=&0  \nonumber \\
\gamma pF_{L}-MF_{RL}-\gamma ^{0}\Phi F_{L} &=&0
\end{eqnarray}

By combining the above equations, one finally arrives to : 
\begin{equation}
(p^{2}-M^{2}-\gamma p\gamma ^{0}\Phi )F_{L}(p)=0  \label{FL}
\end{equation}
\begin{equation}
(p^{2}-M^{2}-\gamma ^{0}p\gamma M\Phi M^{-1})F_{R}(p)=0  \label{FR}
\end{equation}

An important remark must be made. We are here considering a hypothetical
situation where neutrinos had time enough to equilibrate with the background
matter. Under this assumption, right-handed neutrinos can be produced by
different mechanisms, such as spin-flip or pair production. However,
production rates are suppressed by a factor $m_{\nu }/E$ , where $E$ is the
neutrino energy and $m_{\nu }$ its mass. In the astrophysical scenarios we
are considering, the production rate of right-handed neutrinos is small, and
therefore they can be neglected. By this reason, we will concentrate on the
left-handed component $F_{L}(p)$. Consistently with this approximation,
neutrinos will be treated under the extreme relativistic limit $m_{\nu
}/E<<1 $. This will imply that the neutrino field can be considered,
approximately, as consisting on negative-helicity neutrinos and
positive-helicity antineutrinos. For Wigner functions, this is shown in
Appendix B, where the Wigner function will be explicitly calculated in the
interacting case.

The dispersion relation obtained from Eq. (\ref{FL}) can be diagonalized,
and one obtains the well-known expressions for masses and mixing angles in
matter : this is also studied in Appendix B.

\section{Non-equilibrium system. Transport Equation.}

In this section, we investigate the evolution of neutrinos when deviations
from equilibrium situations arise. More precisely, we will consider that
neutrinos are created and propagate through the matter background. This
implies that Eqs. (\ref{cond1}) - (\ref{cond2}) will not be imposed, and
therefore time and spatial variations have to be considered. This represents
an additional difficulty in solving the system of equations for Wigner
functions {\em and} correlation functions. For the moment, we will only
consider a simple case, where correlations are neglected. This will serve us
to investigate the possibilities of Wigner function techniques in deriving
neutrino transport equations, and will allow in the future to study more
complicated frameworks. As we will see in this section, the equations
arising in this context are appropriate to deal with neutrino propagation
and flavor conversion in the Sun.

We return to Eq. (\ref{wigcorra}), and assume that electrons can be locally
characterized by their temperature and chemical potential, in such a way
that Eq. (\ref{wigeqel}) is still valid.

By performing the same procedure as in Eq. (\ref{FL}), we can derive an
equation for $F_{L}(x,p)$, which is now 
\begin{equation}
\lbrack \Box -4(p^{2}-M^{2})-4ip_{\mu }\partial ^{\mu }+2i\Phi (x)\gamma
^{\mu }\gamma ^{0}\partial _{\mu }+4\Phi (x)\gamma ^{\mu }p_{\mu }\gamma
^{0}+2i\gamma ^{\mu }\left( \partial _{\mu }\Phi (x)\right) \gamma
^{0}]F_{L}(x.p)=0  \label{eqmotFL}
\end{equation}

The next step is achieved by decomposing the complete Wigner function $%
F(x.p) $ into the Dirac algebra. By using the chirality projectors, as in
Eq. (\ref{projection}) one can write $F_{L}(x,p)$ under the form 
\begin{equation}
F_{L}(x,p)=\frac{1}{2}(1-\gamma _{5})f_{L\mu }(x,p)\gamma ^{\mu }
\label{structFL}
\end{equation}

where $f_{L\mu }(x,p)$ is a matrix in flavor space\footnote{%
We notice from here that $f_{L\mu }(x,p)$ is an hermitian matrix.}, and
transforms as a Lorentz vector. If the left-handed projector $P_{L}$ is used
again, we see that $f_{L\mu }(x,p)$ can be expressed as 
\begin{equation}
f_{L}^{\mu ,ab}(x,p)=\frac{1}{2}Tr[F_{L}^{ab}(x,p)\gamma ^{\mu }]=\frac{1}{2}%
(2\pi )^{-4}\int d^{4}y\,e^{-ipy}<\widehat{\bar{\nu}}_{L}^{b}(x+\frac{1}{2}%
y)\gamma ^{\mu }\widehat{\nu }_{L}^{a}(x-\frac{1}{2}y)>  \label{52}
\end{equation}

In the latter equation, $\widehat{\nu }_{L}$ is the left-handed component of
the neutrino field. We next analyze the equation of motion obeyed by $%
f_{L\mu }(x,p)$. In order to do this, we substitute Eq. (\ref{structFL})
into Eq. (\ref{eqmotFL}). After some algebra, we obtain the equations 
\begin{eqnarray}
-\frac{1}{4}\Box f_{L}^{0}+(p^{2}-M^{2})f_{L}^{0}-\Phi (x)\left(
p^{0}f_{L}^{0}+\vec{p}\cdot \vec{f}_{L}\right) &&  \nonumber \\
+ip_{\mu }\partial ^{\mu }f_{L}^{0}-\frac{i}{2}\frac{\partial }{\partial t}%
\left( \Phi (x)f_{L}^{0}\right) +\frac{i}{2}\vec{\nabla}\left( \Phi (x)\cdot 
\vec{f}_{L}\right) &=&0  \label{eqmotf0}
\end{eqnarray}
\begin{align}
& -\frac{1}{4}\Box \vec{f}_{L}+(p^{2}-M^{2})\vec{f}_{L}-\Phi (x)\left( p^{0}%
\vec{f}_{L}+\vec{p}f_{L}^{0}\right) +\frac{1}{2}\vec{\nabla}\times \left(
\Phi (x)\vec{f}_{L}\right)  \nonumber \\
& +ip_{\mu }\partial ^{\mu }\vec{f}_{L}-\frac{i}{2}\frac{\partial }{\partial
t}\left( \Phi (x)\vec{f}_{L}\right) +\frac{i}{2}\vec{\nabla}\left( \Phi
(x)f_{L}^{0}\right) +i\Phi (x)\vec{p}\times \vec{f}_{L}=0  \label{eqmotfv}
\end{align}

\ Eqs. (\ref{eqmotf0}) and (\ref{eqmotfv}) are the basic transport equations
to be solved on a general situation, with the help of appropriate boundary
conditions. We will investigate the consequences of this set of equations in
a future work. For the moment, as a test, we will show that, under the
circumstances usually considered when the MSW is studied, we reproduce the
known equations for this effect.

\section{MSW effect}

We consider neutrinos moving along a straight line (for example, the radial
direction of the star). According to this, we assume that $\vec{f}_{L}$ is
parallel to $\vec{p}$. This allows us to write 
\begin{equation}
\vec{f}_{L}(x,p)\equiv \vec{p}f(x,p)
\end{equation}
where $f(x,p)$ is a new function. We also introduce, for convenience, 
\begin{equation}
f_{L}^{0}(x,p)\equiv |\vec{p}|g(x,p)
\end{equation}
Next, we assume that neutrinos are ultrarelativistic, and that the effective
potentials $\widetilde{V}$ and $V_{n}$ in Eq.(\ref{phimatrix}) satisfy 
\begin{equation}
\widetilde{V},V_{n}\ll p^{0}\sim |\vec{p}|\sim 1MeV  \label{relataprox}
\end{equation}
The last condition will concern the characteristic scale of spatial and time
variations of the neutrino distribution function. This scale has a
macroscopic size $R$, at least of the order of 1 Km, or even more (in the
case of the resonant zone in the Sun, for example). In this case, we can
make the following estimate : 
\begin{equation}
\frac{\partial f_{L}^{\mu }}{\partial t}\sim |\vec{\nabla}f_{L}^{\mu }|\sim 
\frac{f_{L}^{\mu }}{R}  \label{macroscopic}
\end{equation}

When $R=1$ Km, then $1/R\sim 10^{-16}MeV$. If we are interested on
variations of the distribution function on the scale $R$, then, together
with hypothesis Eq. (\ref{relataprox}) we can simplify Eqs. (\ref{eqmotf0})
and (\ref{eqmotfv}) to give 
\begin{eqnarray}
\left[ p^{2}-M^{2}-(p^{0}+|\vec{p}|)\Phi (x)\right] (f+g)+ip_{\mu }\partial
^{\mu }(f+g) &=&0  \nonumber \\
\left[ p^{2}-M^{2}-(p^{0}-|\vec{p}|)\Phi (x)\right] (f-g)+ip_{\mu }\partial
^{\mu }(f-g) &=&0  \label{sistfg}
\end{eqnarray}
In the latter equations, $f(x,p)$ and $g(x,p)$ are {\em Hermitian} matrices
(in flavor space). We can perform a {\em local} transformation for each of
them in such a way that both become diagonal (and therefore real). In this
way, one can easily check that two possibilities are open for the system Eq.
(\ref{sistfg}) to have non-trivial solutions. The two possibilities are 
\begin{equation}
\det \left[ p^{2}-M^{2}-(p^{0}\pm |\vec{p}|)\Phi (x)\right] =0
\label{dispertwog}
\end{equation}
We recognize in Eq. (\ref{dispertwog}) the dispersion relations for
neutrinos and antineutrinos, as described in Appendix B within the Hartree
equilibrium hypothesis. However, quantities in the latter equation depend on
the coordinate $x$ . This means that the neutrino mass eigenstates can be
obtained locally from the Hartree approximation (which gives the same as the
MSW effect). We then have, for neutrinos, $\det \left[ p^{2}-M^{2}-(p^{0}+|%
\vec{p}|)\Phi (x)\right] =0$. This implies the condition $f(x,p)=g(x,p)$.
Therefore, $f_{L}^{0}(x,p)=|\vec{p}|f(x,p)$ $\simeq p^{0}f(x,p)$ , and we
conclude that 
\begin{equation}
f_{L}^{\mu }(x,p)\simeq p^{\mu }f(x,p)
\end{equation}
within the same approximation. Finally, we have for Eq. (\ref{structFL}) 
\begin{equation}
F_{L}(x,p)=\frac{1}{2}(1-\gamma _{5})p_{\mu }\gamma ^{\mu }f(x,p)
\end{equation}

where $f(x,p)$ , for ultra-relativistic neutrinos, obeys the following
equation of motion 
\begin{equation}
\left[ p^{2}-M^{2}-2\Phi (x)p^{0}\right] f+ip_{\mu }\partial ^{\mu }f=0
\end{equation}

We have discussed above the possibility of making a local transformation
which diagonalizes $f(x,p)$. We have exploited the fact that, under these
circumstances, it becomes a real matrix. However, from the physical point of
view, it is more convenient to introduce a different local transformation,
in such a way that the factor inside the brackets in the latter equation
becomes diagonal, i.e. we consider an unitary transformation given by the
matrix $U_{M}(x)$

\begin{equation}
\widetilde{f}=U_{M}^{\dagger }fU_{M}  \label{transfdiag}
\end{equation}
such that 
\begin{equation}
U_{M}^{\dagger }(x)[(p^{2}-M^{2})-2\Phi (x)p^{0}]U_{M}(x)=p^{2}-\widetilde{M}%
^{2}(x)
\end{equation}

where $\widetilde{M}^{2}(x)\equiv U_{M}^{\dagger }(x)\left[ M^{2}+2\Phi
(x)p^{0}\right] U_{M}(x)$ is a diagonal matrix, which contains the local
mass eigenvalues \footnote{%
We use a tilde to represent magnitudes in the interacting eigenstates basis,
as mentioned in Section 2.}: 
\begin{equation}
\widetilde{M}^{2}(x)=\left( 
\begin{array}{cc}
\widetilde{M}_{1}^{2}(x) & 0 \\ 
0 & \widetilde{M}_{2}^{2}(x)
\end{array}
\right)  \label{massmatrix}
\end{equation}

On the other hand, one finds 
\begin{equation}
U_{M}^{\dagger }(x)\left( i\partial ^{\mu }p_{\mu }f(x,p)\right)
U_{M}(x)=i\partial ^{\mu }p_{\mu }\tilde{f}(x,p)-[\tilde{f}%
(x,p),U_{M}^{\dagger }(x)i\partial ^{\mu }p_{\mu }U_{M}(x)]  \label{71}
\end{equation}
Thus the equation of motion reads 
\begin{equation}
i\partial ^{\mu }p_{\mu }\tilde{f}(x,p)+(p^{2}-\widetilde{M}^{2}(x))\tilde{f}%
(x,p)-i[\tilde{f}(x,p),U_{M}^{\dagger }(x)p_{\mu }\partial ^{\mu }U_{M}(x)]=0
\label{motftilde}
\end{equation}

Let us write explicitly the matrices $\tilde{f}(x,p)$ and $U_{M}(x)$ by
defining

\begin{equation}
U_{M}(x)\equiv \left( 
\begin{array}{cc}
\cos \theta _{M}(x) & -\sin \theta _{M}(x) \\ 
\sin \theta _{M}(x) & \cos \theta _{M}(x)
\end{array}
\right)  \label{transfmatrix}
\end{equation}

\begin{equation}
\tilde{f}(x,p)\equiv \left( 
\begin{array}{cc}
\tilde{f}^{11}(x,p) & \tilde{f}^{12}(x,p) \\ 
\tilde{f}^{21}(x,p) & \tilde{f}^{22}(x,p)
\end{array}
\right)  \label{explicitftilde}
\end{equation}

where the functions $\tilde{f}^{11}(x,p)$ and $\tilde{f}^{22}(x,p)$\ are
real, while $\tilde{f}^{12}(x,p)$\ and $\tilde{f}^{21}(x,p)$\ are the
complex conjugate of one another. After substituting Eqs. (\ref{massmatrix}%
),(\ref{transfmatrix}) and (\ref{explicitftilde}) into Eq. (\ref{motftilde}%
), one obtains 
\begin{equation}
ip^{\mu }\partial _{\mu }\widetilde{f}^{11}(x,p)+\left( p^{2}-\widetilde{M}%
_{1}^{2}(x)\right) \widetilde{f}^{11}(x,p)-ip^{\mu }\partial _{\mu }\theta
_{M}(x)\left( \widetilde{f}^{12}(x,p)+\widetilde{f}^{21}(x,p)\right) =0
\label{f11}
\end{equation}
\begin{equation}
ip^{\mu }\partial _{\mu }\widetilde{f}^{12}(x,p)+\left( p^{2}-\widetilde{M}%
_{1}^{2}(x)\right) \widetilde{f}^{12}(x,p)+ip^{\mu }\partial _{\mu }\theta
_{M}(x)\left( \widetilde{f}^{11}(x,p)-\widetilde{f}^{22}(x,p)\right) =0
\label{f12}
\end{equation}
\begin{equation}
ip^{\mu }\partial _{\mu }\widetilde{f}^{21}(x,p)+\left( p^{2}-\widetilde{M}%
_{2}^{2}(x)\right) \widetilde{f}^{21}(x,p)+ip^{\mu }\partial _{\mu }\theta
_{M}(x)\left( \widetilde{f}^{11}(x,p)-\widetilde{f}^{22}(x,p)\right) =0
\label{f21}
\end{equation}
\begin{equation}
ip^{\mu }\partial _{\mu }\widetilde{f}^{22}(x,p)+\left( p^{2}-\widetilde{M}%
_{2}^{2}(x)\right) \widetilde{f}^{22}(x,p)+ip^{\mu }\partial _{\mu }\theta
_{M}(x)\left( \widetilde{f}^{12}(x,p)+\widetilde{f}^{21}(x,p)\right) =0
\label{f22}
\end{equation}

A remark is in order. As can be seen from the above equations, it is not
possible, in a general situation out of equilibrium, to perform a
transformation that makes both the neutrino Wigner function {\em and} the
mass matrix diagonal. However, local mass eigenstates can be used as a
useful physical basis to simplify the equations, as done in this section.

We can write the latter system of equations in a more familiar way. By
taking the real part on the first and last one, we readily arrive to the
conditions 
\begin{eqnarray}
\left[ p^{2}-\widetilde{M}_{1}^{2}(x)\right] \widetilde{f}^{11}(x,p)
&=&0\,\,\,  \nonumber \\
\left[ p^{2}-\widetilde{M}_{2}^{2}(x)\right] \widetilde{f}^{22}(x,p)
&=&0\,\,\,
\end{eqnarray}

which imply that $\widetilde{f}^{11}(x,p)$ ($\widetilde{f}^{22}(x,p)$) is
non-vanishing only when $p^{2}=\widetilde{M}_{1}^{2}(x)$ ( $p^{2}=\widetilde{%
M}_{2}^{2}(x)$). Similarly, from the two remaining equations it can be
easily deduced that the functions $\widetilde{f}^{12}(x,p)$ and $\widetilde{f%
}^{21}(x,p)$ have to vanish unless the condition $p^{2}=\frac{1}{2}\left[ 
\widetilde{M}_{1}^{2}(x)+\widetilde{M}_{2}^{2}(x)\right] $ is fulfilled.
Therefore, in the equations of motion for these functions we can substitute 
\begin{eqnarray}
p^{2}-\widetilde{M}_{1}^{2}(x) &\rightarrow &\frac{1}{2}\Delta (x)  \nonumber
\\
p^{2}-\widetilde{M}_{2}^{2}(x) &\rightarrow &-\frac{1}{2}\Delta (x)
\end{eqnarray}

where $\Delta (x)\equiv \widetilde{M}_{2}^{2}(x)-\widetilde{M}_{1}^{2}(x)$
is the in-medium neutrino mass difference. Furthermore, in the
ultra-relativistic case, we can approximate the operator $p^{\mu }\partial
_{\mu }$ in {\em all} the Eqs. (\ref{f11})-(\ref{f22}) by 
\begin{equation}
p^{\mu }\partial _{\mu }\simeq |\vec{p}|\frac{\partial }{\partial t}+|\vec{p}%
|\frac{\partial }{\partial x}=|\vec{p}|D
\end{equation}

We have defined $D=\frac{\partial }{\partial t}+\frac{\partial }{\partial x}$%
. After some manipulations, Eqs. (\ref{f11})-(\ref{f22}) can be cast under
the simple matrix form 
\begin{equation}
iD\widetilde{f}(x,t)=[\widetilde{f}(x,t),\widetilde{H}(x)]  \label{mswftilde}
\end{equation}

Here, 
\begin{equation}
\widetilde{H}(x)=\left( 
\begin{array}{cc}
\frac{\Delta (x)}{4\vec{p}|} & -i\theta _{M}^{\prime }(x) \\ 
i\theta _{M}^{\prime }(x) & -\frac{\Delta (x)}{4|\vec{p}|}
\end{array}
\right)  \label{53}
\end{equation}

is the {\em effective Hamiltonian} in the mass eigenstates basis, and $%
\theta _{M}^{\prime }(x)=\frac{d}{dx}\theta _{M}(x)$. Eq. (\ref{mswftilde})
can also be written in the flavor basis, if we undo the transformation
introduced in Eq. (\ref{transfdiag}), with the definitions 
\begin{equation}
f(x,t)\equiv \left( 
\begin{array}{cc}
f^{ee}(x,t) & f^{e\mu }(x,t) \\ 
f^{\mu e}(x,t) & f^{\mu \mu }(x,t)
\end{array}
\right) =U_{M}(x)\widetilde{f}(x,t)U_{M}^{\dagger }(x)  \label{55}
\end{equation}

After a straightforward calculation, we arrive to the equation 
\begin{equation}
iDf(x,t)=[f(x,t),H(x)]  \label{mswf}
\end{equation}

with 
\begin{equation}
H(x)=\left( 
\begin{array}{cc}
\alpha (x) & \beta \\ 
\beta & -\alpha (x)
\end{array}
\right)  \label{63}
\end{equation}

is the Hamiltonian in the flavor basis, and we introduced the notations 
\begin{eqnarray}
\alpha (x) &=&(\Delta _{0}\cos 2\theta -A(x))/4|\vec{p}|  \nonumber \\
\beta &=&\Delta _{0}\sin 2\theta /4|\vec{p}|
\end{eqnarray}

where $A(x)=2|\vec{p}|\sqrt{2}G_{F}\,n(x)$ is the induced mass due to
charged currents, $\Delta _{0}=m_{2}^{2}-m_{1}^{2}$ is the vacuum neutrino
mass difference and $\theta $ the vacuum mixing angle.

From the system of equations (\ref{mswf}), we first obtain that 
\begin{equation}
iD\left[ f^{ee}(x,t)+f^{\mu \mu }(x,t)\right] =0\,\,  \label{68}
\end{equation}

which implies that the total number of electron plus muon neutrinos is
conserved during the propagation, as expected. We can then write $%
f^{ee}(x,t)+f^{\mu \mu }(x,t)=K$ , where $K$ is constant during the
propagation : $DK=0$. By manipulating the equations, one can also derive a
third-order differential equation for $f^{ee}(x,t)$, which reads as 
\begin{equation}
\alpha D^{3}f^{ee}(x,t)-\alpha ^{\prime }D^{2}f^{ee}(x,t)+4\alpha (\alpha
^{2}+\beta ^{2})Df^{ee}(x,t)-4\alpha ^{\prime }\beta ^{2}\left( f^{ee}(x,t)-%
\frac{K}{2}\right) =0  \label{74}
\end{equation}
where $\alpha ^{\prime }=\frac{d}{dx}\alpha $. \ The latter equation
coincides with the one derived by Mikheyev and Smirnov \cite{MS86b} to
describe the evolution of the survival probability of an electron neutrino
in a non-constant density medium. This agrees with our interpretation of $%
f^{ee}(x,t)$ as giving the distribution function (proportional to the number
density for a given momentum) of electron neutrinos, and analogously for $%
f^{\mu \mu }(x,t)$ as the distribution function of muon neutrinos. All known
results for the MSW effect within the situation considered here can be
reproduced from the above equations.

Eq. (\ref{mswf}) defines the evolution of flavor distribution functions, and
can be compared to the corresponding results derived from other treatments.
As we mentioned in the introduction, there are (to our knowledge) two
different approaches to neutrino propagation in dense media which make use
of some kind of distribution functions. Both methods are based on
perturbation techniques, assuming that the Hamiltonian can be separated into
two terms : $H=H_{0}+H_{int}$, with $H_{int}$ considered as a small
perturbation. They both arrive to an equation with a term which is second
order in $G_{F}$ and corresponds to a non-trivial (i.e., non-diagonal in
flavor) Boltzmann collision integral of neutrinos interacting with other
particles in the background. This second-order term is absent in our
approach, at least when correlations are neglected. Within this
approximation, and neglecting small derivative terms\footnote{%
As discussed in \cite{Ru90} , these terms should only kept in the equations
when strong inhomogeneities of size $L$ are present, such that $|\vec{p}%
|L\sim 1$. See also\cite{Se87} and \cite{OSeSmo94}.}, \ the evolution
equations of Ref. \cite{Ru90} coincides with Eq. (\ref{mswf}). A similar
comparison can be made using the results derived in \cite{RSiSt93},\cite
{RS93}.

\section{Conclusions}

In this paper we have studied propagation of two neutrino species in dense
media with flavor oscillations. We used an scheme which is based on the
introduction of relativistic Wigner functions and correlation functions. Our
aim is to develop relativistic kinetic equations, and to analyze the
possibility that such approach will correctly describe both the relativistic
dispersion relations of in-medium neutrinos, as well as the appropriate
evolution equations for neutrino distribution functions. This allows us to
treat in-medium neutrino masses and kinetic equations on an equal foot. We
considered the medium as consisting on electrons, although other
constituents can be incorporated in a straightforward way.

By writing the equations of motion for Wigner operators and taking
statistical averages, one arrives to an infinite chain of equations of the
BBGKY hierarchy-type. This infinite chain has to be broken at some level in
order to obtain a solution, and we have considered here the lowest-level of
approximation, which consists in neglecting correlations (the so-called
Hartree approximation). We have first examined the situation in equilibrium,
which reproduces the well-known results for relativistic dispersion
relations \cite{NR88}. Next, we assumed that the electron background is in
equilibrium, although its density needs not to be constant, and neutrinos
propagate out of equilibrium. This gives rise to a set of kinetic equations
which have to be solved, in a general situation, by taking the appropriate
boundary conditions.

In order to obtain some insight into the above equations, we have considered
with some detail the usual MSW scenario, in which neutrinos are relativistic
and the scale of space-time variations in distribution functions have
macroscopic values (such as flavor conversion in the Sun or newly-born
neutron stars). In this case, our system of equations reduce to the results
obtained by other authors with the use of perturbation techniques \cite{Ru90}%
,\cite{RSiSt93} when only the first-order correction is considered. However,
none of these methods has been showed to incorporate the correct neutrino
dispersion relations.

Now the question which arises is wether the inclusion of correlations into
our scheme will lead to the same results as in the previous references, also
for the second-order terms. Indeed, this seems to be the case, since
correlations turn out to be proportional to the coupling constant $G_{F}$,
and substitution into the BBGKY hierarchy will give corrections of the order 
$G_{F}^{2}$, as one can see from Eqs. (\ref{wigcorra}) and (\ref{wigcorrb}).
A more detailed study of the kinetic equations developed here will elucidate
this question and, perhaps, give rise to new phenomena in the physics of
neutrinos in dense media. This will be the subject of a future work.

\appendix

\section{Free system with one and two generations}

In this appendix we give explicit formulae for the non-interacting neutrino
Wigner functions. We first consider the case with only one generation, and
we will generalize these results to the case of two neutrino generations
with a non-diagonal mass matrix. Neutrinos are assumed to be in equilibrium.
Of course, equilibrium can not be reached in the non-interacting case, but
one can imagine a situation where a very weak interaction is added in order
for the system to reach equilibrium. This hypothetical interaction can then
be turned off without changing the equilibrium properties of the system.

We start with one generation of free neutrinos in equilibrium. The Wigner
function then verifies the equations : 
\begin{eqnarray}
&(&\gamma p-m)F^{(\nu )}(p)=0  \nonumber \\
&&F^{(\nu )}(p)(\gamma p-m)=0
\end{eqnarray}

where $m$ is the neutrino mass and $F^{(\nu )}(p)$ the equilibrium Wigner
function. As in previous cases, the above equations are the Hermitian
conjugate of one another. We next multiply the first one by $(\gamma p+m)$,
which gives 
\begin{equation}
(p^{2}-m^{2})F^{(\nu )}(p)=0  \label{327}
\end{equation}

and this implies that the Wigner function vanishes whenever the condition $%
p^{2}-m^{2}=0$ is not fulfilled. We introduce the Grand Canonical density
matrix operator 
\begin{equation}
\widehat{\rho }=Z^{-1}e^{-\beta \,(\widehat{H}-\mu \,\widehat{L})};\qquad
Z=Tr\,e^{-\beta \,(\widehat{H}-\mu \,\widehat{L})}  \label{329}
\end{equation}

where $\widehat{H}$\ is the Hamiltonian, $\widehat{L}$\ the lepton number
operator and $\mu $\ the neutrino chemical potential.

We obtain, after standard quantization of the neutrino field in the
helicity-states basis: 
\begin{eqnarray}
&&F_{ij}^{(\nu )}(p)=(2\pi )^{-4}\int d^{4}y\,e^{-ipy}<\widehat{\bar{\nu}}%
_{j}(x+y/2)\widehat{\nu }_{i}(x-y/2)>=  \nonumber \\
&=&(2\pi )^{-4}\int d^{4}y\,e^{-ipy}(2\pi )^{-3}\int d^{3}\vec{k}%
\,\sum_{\lambda =\pm }[<\widehat{N}_{\lambda }(\vec{k})>\bar{u}_{j}^{\lambda
}(\vec{k})u_{i}^{\lambda }(\vec{k})e^{iky}-<\widehat{\bar{N}}_{\lambda }(%
\vec{k})>\bar{v}_{j}^{\lambda }(\vec{k})v_{i}^{\lambda }(\vec{k})e^{-iky}] 
\nonumber \\
&=&(2\pi )^{-3}\int d^{3}\vec{k}\,\sum_{\lambda =\pm }[<\widehat{N}_{\lambda
}(\vec{k})>\bar{u}_{j}^{\lambda }(\vec{k})u_{i}^{\lambda }(\vec{k})\delta
(p-k)-<\widehat{\bar{N}}_{\lambda }(\vec{k})>\bar{v}_{j}^{\lambda }(\vec{k}%
)v_{i}^{\lambda }(\vec{k})\delta (p+k)]  \label{wigfreeav1}
\end{eqnarray}
where $\widehat{N}_{\lambda }(\vec{k})$ ( $\widehat{\bar{N}}_{\lambda }(\vec{%
k})$) is the number operator for neutrinos (antineutrinos) with momentum $%
\vec{k}$ and helicity $\lambda \footnote{%
Operators will be considered in normal order.}$. The notations $u^{\lambda }(%
\vec{k}),v^{\lambda }(\vec{k}),...$ for spinors and $\widehat{a}^{\dagger }(%
\vec{k}),\widehat{a}_{\lambda }(\vec{k}),...$ for creation and destruction
operators have their usual meaning. The above averages can be calculated
using standard techniques, giving : 
\begin{equation}
<\hat{N}_{\lambda }(\vec{k})>=Tr(\widehat{\rho }\widehat{a}_{\lambda
}^{\dagger }(\vec{k})\widehat{a}_{\lambda }(\vec{k}))=\frac{1}{e^{\beta
\,(E_{k}-\mu )}+1}\equiv f(k)  \label{331}
\end{equation}
\begin{equation}
<\widehat{\bar{N}}_{\lambda }(\vec{k})>=Tr(\widehat{\rho }\widehat{b}%
_{\lambda }^{\dagger }(\vec{k})\widehat{b}_{\lambda }(\vec{k}))=\frac{1}{%
e^{\beta \,(E_{k}+\mu )}+1}\equiv \bar{f}(k)  \label{332}
\end{equation}
with $E_{k\text{ }}=\sqrt{m^{2}+\vec{k}^{2}}$. Let us define the matrices 
\begin{eqnarray}
\Sigma _{ij}^{\lambda }(\vec{k}) &=&\bar{u}_{j}^{\lambda }(\vec{k}%
)u_{i}^{\lambda }(\vec{k})  \nonumber \\
\bar{\Sigma}^{\lambda }(\vec{k}) &=&-\bar{v}_{j}^{\lambda }(-\vec{k}%
)v_{i}^{\lambda }(-\vec{k})
\end{eqnarray}

More explicitly, if we introduce a coordinate system in such a way that $%
\vec{k}=k(\sin \theta \cos \phi ,\sin \theta \sin \phi ,\cos \theta )$ one
has 
\begin{equation}
\Sigma ^{+}(\vec{k})=\frac{1}{2E_{k}}\left( 
\begin{array}{cc}
m & -E_{k}-k \\ 
-E_{k}+k & m
\end{array}
\right) \mu (\vec{k})  \label{334}
\end{equation}
\begin{equation}
\Sigma ^{-}(\vec{k})=\frac{1}{2E_{k}}\left( 
\begin{array}{cc}
m & -E_{k}+k \\ 
-E_{k}-k & m
\end{array}
\right) \rho (\vec{k})  \label{335}
\end{equation}
\begin{equation}
\bar{\Sigma}^{+}(\vec{k})=\frac{1}{2E_{k\text{ }}}\left( 
\begin{array}{cc}
m & E_{k\text{ }}-k \\ 
E_{k\text{ }}+k & m
\end{array}
\right) \mu (\vec{k})  \label{346}
\end{equation}
\begin{equation}
\bar{\Sigma}^{-}(\vec{k})=\frac{1}{2E_{k}}\left( 
\begin{array}{cc}
m & E_{k\text{ }}+k \\ 
E_{k\text{ }}-k & m
\end{array}
\right) \rho (\vec{k})  \label{347}
\end{equation}

where 
\begin{equation}
\mu (\vec{k})=\left( 
\begin{array}{cc}
\cos ^{2}\frac{\theta }{2} & \frac{1}{2}\sin \theta \,e^{-i\phi } \\ 
\frac{1}{2}\sin \theta \,e^{i\phi } & \sin ^{2}\frac{\theta }{2}
\end{array}
\right)  \label{338}
\end{equation}

and $\rho (\vec{k})=\mu (-\vec{k})$. By substituting into Eq. (\ref
{wigfreeav1}) one obtains, after some algebra 
\begin{equation}
F^{(\nu )}(p)=(2\pi )^{-3}2E_{p}\delta (p^{2}-M^{2})\sum_{\lambda =\pm
}[\theta (p^{0})f(p)\Sigma ^{\lambda }(\vec{p})+\theta (-p^{0})\bar{f}(p)%
\bar{\Sigma}^{\lambda }(\vec{p})]  \label{349}
\end{equation}

By performing the sum over helicities, we finally arrive to the expression 
\begin{equation}
F^{(\nu )}(p)=(\gamma p+m)f_{W}(p)  \label{359}
\end{equation}

where the scalar function $f_{W}(p)$ is given by 
\begin{equation}
f_{W}(p)=\frac{1}{4m}Tr\left[ F^{(\nu )}(p)\right] =(2\pi )^{-3}\delta
(p^{2}-m^{2})\left[ \theta (p^{0})f(p)+\theta (-p^{0})\bar{f}(p)\right]
\label{360}
\end{equation}

The generalization of the above formulae to more than one neutrino flavor is
done by using mass eigenstates as an intermediate step. As defined in
Section 2, we use a prime to represent such states. In the case of
non-interacting neutrinos, they arise from diagonalization of the free mass
matrix $M$. The equation of motion for the neutrino Wigner function is now,
in the flavor basis 
\begin{equation}
(\gamma p-M)F^{(\nu )}(p)=0  \label{362}
\end{equation}

The free mass eigenstates are related to the flavor states through a matrix 
\begin{equation}
U\equiv \left( 
\begin{array}{cc}
\cos \theta & -\sin \theta \\ 
\sin \theta & \cos \theta
\end{array}
\right)
\end{equation}

in the way 
\begin{equation}
\widehat{\nu }(x)=U\widehat{\nu }^{\prime }(x)
\end{equation}

and one has 
\begin{equation}
M=UM^{\prime }U^{\dagger }
\end{equation}

where $M^{\prime }$ is the eigenvalues matrix 
\begin{equation}
M^{\prime }=diag(m_{1},m_{2})
\end{equation}

The Wigner function in the mass eigenstates basis, $F^{\prime \,(\nu )\,}(p)$
is related to $F^{(\nu )}(p)$ trough 
\begin{equation}
F^{(\nu )}(p)=UF^{\prime \,(\nu )\,}(p)U^{\dagger }  \label{FprimetoF}
\end{equation}

and will obey the diagonal equation 
\begin{equation}
(p^{2}-M^{\prime \,2})F^{\prime (\nu )}(p)=0  \label{twogenfree}
\end{equation}

By quantizing the massive fields $\widehat{\nu }^{\prime a}(x)$ ($a=1,2$ for
two generations) we obtain the result 
\begin{eqnarray}
&&F_{ij}^{\prime \,(\nu )ab}(p)=(2\pi )^{-4}\int d^{4}y\,e^{-ipy}<\widehat{%
\bar{\nu}}_{j}^{\prime b}(x+y/2)\widehat{\nu }_{i}^{\prime a}(x-y/2)>= 
\nonumber \\
&=&(2\pi )^{-3}\int d^{3}\vec{k}\,\sum_{\lambda =\pm }[<\widehat{N}_{\lambda
}^{a}(\vec{k})>\bar{u}_{j}^{a,\lambda }(\vec{k})u_{i}^{a,\lambda }(\vec{k}%
)\delta (p-k)-  \nonumber \\
&<&\widehat{\bar{N}}_{\lambda }^{a}(\vec{k})>\bar{v}_{j}^{a,\lambda }(\vec{k}%
)v_{i}^{a,\lambda }(\vec{k})\delta (p+k)]\delta _{a,b}\qquad a,b=1,2
\end{eqnarray}

with similar notations as before. The superscript $a$ has been added to
label different mass eigenstates. In this basis, then, the Wigner function
is diagonal, as expected from Eq. (\ref{twogenfree}). The Wigner functions
appearing in the latter equation are given by 
\begin{eqnarray}
F^{\prime \,(\nu )11}(p) &=&(2\pi )^{-3}\delta (p^{2}-m_{1}^{2})\left[
\theta (p^{0})f_{1}(p)+\theta (-p^{0})\bar{f}_{1}(p)\right] (\gamma p+m_{1})
\nonumber \\
F^{\prime \,(\nu )22}(p) &=&(2\pi )^{-3}\delta (p^{2}-m_{2}^{2})\left[
\theta (p^{0})f_{2}(p)+\theta (-p^{0})\bar{f}_{2}(p)\right] (\gamma p+m_{2})
\label{uy}
\end{eqnarray}

and we introduced the notations 
\begin{eqnarray}
f_{a}(p) &=&\frac{1}{e^{\beta (E_{p}^{a}-\mu )}+1}  \label{f2gplus} \\
\bar{f}_{a}(p) &=&\frac{1}{e^{\beta (E_{p}^{a}+\mu )}+1}  \label{f2gminus}
\end{eqnarray}

($a=1,2$) with $E_{p}^{a}\equiv \sqrt{m_{a}^{2}+\vec{p}^{2}}$. We notice
that the chemical potential is the same for both generations of neutrinos in
Eq. (\ref{f2gplus}). This is so because, in our model, there is only a
conserved current : the electron plus muon lepton number. A similar comment
can be made regarding antineutrinos, Eq. (\ref{f2gminus}). The Wigner
function in the flavor basis is then obtained from Eq. (\ref{FprimetoF}).

\section{Neutrino masses and Wigner functions in the Hartree approximation}

We now discuss with some detail the dispersion relations and neutrino masses
which arise from the equations of motion of the Wigner function within the
Hartree approximation. In order to simplify the notations as much as
possible, and to obtain insight into the problem as well, we start with one
neutrino flavor (let us consider electron neutrinos). Then, Eq. (\ref
{wigeqneu}) takes the simpler form 
\begin{equation}
\lbrack \gamma p-m-\gamma ^{0}\frac{1}{2}(1-\gamma ^{5})\widetilde{V}%
]F^{(\nu )}(p)=0  \label{eqwigoneg}
\end{equation}

Dispersion relations are the necessary conditions for Eq. (\ref{eqwigoneg})
to have solutions other than the trivial one $F^{(\nu )}(p)=0$. One can
compute the determinant for this equation by setting $p=(p_{0},0,0,p_{z})$.
The result then is 
\begin{equation}
\det [\gamma p-m-\gamma ^{0}\frac{1}{2}(1-\gamma ^{5})\widetilde{V}]=\left[
p_{0}^{2}-p_{z}^{2}-m^{2}-\left( p_{z}+p_{0}\right) \widetilde{V}\right] %
\left[ p_{0}^{2}-p_{z}^{2}-m^{2}+\left( p_{z}-p_{0}\right) \widetilde{V}%
\right]
\end{equation}

By equating to zero this determinant, and solving for $p_{0}$, one obtains
four solutions : $p_{0}=E_{+}$ , $p_{0}=E_{-}$, $p_{0}=-\bar{E}_{+}$ and $%
p_{0}=-\bar{E}_{-}$, where 
\begin{equation}
E_{+}\equiv \frac{\widetilde{V}}{2}+\sqrt{(\frac{\widetilde{V}}{2}%
-p_{z})^{2}+m^{2}}  \label{110}
\end{equation}
\begin{equation}
E_{-}\equiv \frac{\widetilde{V}}{2}+\sqrt{(\frac{\widetilde{V}}{2}%
+p_{z})^{2}+m^{2}}  \label{111}
\end{equation}
\begin{equation}
\bar{E}_{+}\equiv -\frac{\widetilde{V}}{2}+\sqrt{(\frac{\widetilde{V}}{2}%
-p_{z})^{2}+m^{2}}  \label{112}
\end{equation}
\begin{equation}
\bar{E}_{-}\equiv -\frac{\widetilde{V}}{2}+\sqrt{(\frac{\widetilde{V}}{2}%
+p_{z})^{2}+m^{2}}  \label{113}
\end{equation}

In order to identify the above results, we start from Eq. (\ref{eqwigoneg})
and use the projection operators to obtain an equation for $F_{L}^{(\nu
)}(p) $, as in section 3. One then arrives to 
\begin{equation}
(p^{2}-m^{2}-\gamma p\gamma ^{0}\widetilde{V})F_{L}^{(\nu )}(p)=0
\label{398}
\end{equation}

We take into account the following identity 
\begin{equation}
\gamma p\gamma ^{0}=\left( 
\begin{array}{cc}
p^{0}+\vec{p}\,\cdot \vec{\sigma} & 0 \\ 
0 & p^{0}-\vec{p}\,\cdot \vec{\sigma}
\end{array}
\right)  \label{3103}
\end{equation}

This means that the positive-helicity component $F_{L}^{+}(p)$ has to obey
the relationship 
\begin{equation}
\lbrack p^{2}-m^{2}-\widetilde{V}(p_{0}-|\vec{p}|)]F_{L}^{+}(p)=0
\label{104}
\end{equation}

The solutions for the dispersion relation $p^{2}-m^{2}-\widetilde{V}(p_{0}-|%
\vec{p}|)=0$ are given by $p_{0}=E_{+}$ and $p_{0}=-\bar{E}_{+}$. One can
also make an analogous study for $F_{L}^{-}(p)$, which gives $p_{0}=E_{-}$
and $p_{0}=-\bar{E}_{-}$ as possible solutions. This allows us to interpret
Eqs. (\ref{110})-(\ref{113}) as corresponding to neutrinos ($E$) and
antineutrinos ($\bar{E}$) of positive and negative helicity (subindexes $+$
and $-$, respectively). One can check that these solutions, obtained from
the Hartree approximation, coincide with the ones obtained from the poles of
the neutrino propagator at the one-loop level (see, for example, \cite{NR88}%
).

We now concentrate on ultra-relativistic neutrinos, and consider densities
as occur in normal or compact stars. Under these conditions, we can neglect
terms such as $\widetilde{V}|\overrightarrow{p}|$ and $m^{2}$ by comparison
to 
%TCIMACRO{\TEXTsymbol{\vert}}%
%BeginExpansion
\mbox{$\vert$}%
%EndExpansion
$\overrightarrow{p}|^{2}$ and, in this way, the dispersion relations give us
the neutrinos effective masses for each degree of freedom, which are 
\begin{equation}
M_{+}^{2}=m^{2}  \label{mef1}
\end{equation}

\begin{equation}
M_{-}^{2}=m^{2}+2\widetilde{V}|\overrightarrow{p}|  \label{mef2}
\end{equation}
\begin{equation}
\overline{M}_{+}^{2}=m^{2}-2\widetilde{V}|\overrightarrow{p}|  \label{mef3}
\end{equation}

\begin{equation}
\overline{M}_{-}^{2}=m^{2}  \label{mef4}
\end{equation}
We used the same notations as for the energies. In these equations we see
that positive polarization neutrinos and negative polarization antineutrinos
behave, approximately, as free particles.

The above results can be generalized to the case of two neutrino flavors in
a straightforward way. Let us start from Eq. (\ref{FL}) and transform it to
the mass eigenstates basis in vacuum, as discussed in Appendix A for the
non-interacting case. We then arrive to the following equation 
\begin{equation}
(p^{2}-M^{\prime \,2}-\gamma p\gamma ^{0}\Phi ^{\prime })F_{L}^{\prime
\,(\nu )}(p)=0  \label{119}
\end{equation}

Following the discussion we made above for one generation, we readily obtain
the equation for the negative-helicity component (we replace $p_{0}$ to $%
E_{-}$ for neutrinos) 
\begin{equation}
(p^{2}-M^{\prime \,2}-(E_{-}+|\vec{p}|)\Phi ^{\prime })F_{L}^{-\prime \,(\nu
)}(p)=0  \label{121}
\end{equation}

The latter equation suggests us to define an {\em effective mass matrix }$%
{\cal M}_{-}$ as follows 
\begin{equation}
{\cal M}_{-}^{2}\equiv M^{\prime \,2}+(E_{-}+|\vec{p}|)\Phi ^{\prime }
\label{125}
\end{equation}
If flavor mixing exists, this matrix is non-diagonal. More explicitly, we
find 
\begin{equation}
{\cal M}_{-}^{2}=\left( 
\begin{array}{cc}
(V_{n}+V\cos ^{2}\theta )(E_{-}+|\vec{p}|)+m_{1}^{2} & -V\sin \theta \cos
\theta (E_{-}+|\vec{p}|) \\ 
-V\sin \theta \cos \theta (E_{-}+|\vec{p}|) & (V_{n}+V\sin ^{2}\theta
)(E_{-}+|\vec{p}|)+m_{2}^{2}
\end{array}
\right)  \label{effectivem}
\end{equation}

The corresponding dispersion relation is then 
\begin{equation}
\det \left( p^{2}-{\cal M}_{-}^{2}\right) =0  \label{dispermm}
\end{equation}

Eqs. (\ref{125}) and (\ref{dispermm})can be used to obtain the {\em exact}\
energy levels of neutrinos (and antineutrinos) as a function of the neutrino
momentum. In the ultra-relativistic limit, one can approximate 
\begin{equation}
{\cal M}_{-}^{2}\simeq \left( 
\begin{array}{cc}
A_{n} & 0 \\ 
0 & A_{n}
\end{array}
\right) +\left( 
\begin{array}{cc}
A\cos ^{2}\theta +m_{1}^{2} & -A\sin \theta \cos \theta \\ 
-A\sin \theta \cos \theta & A\sin ^{2}\theta +m_{2}^{2}
\end{array}
\right)  \label{relativism}
\end{equation}

here, $A\equiv 2|\vec{p}|V$ is the induced squared mass due to charged
currents and $A_{n}\equiv 2|\vec{p}|V_{n}$ the analogous magnitude for
neutral currents. We first consider the first term on the second hand of Eq.
(\ref{relativism}) in order to find the eigenvalues of ${\cal M}_{-}^{2}$
(following the notations introduced in Section 2, we represent the diagonal
form of ${\cal M}_{-}$ by $\tilde{M}$\ ). Since neutral currents are
diagonal in flavor, they can be added at the end. After some trivial
algebra, the mass eigenvalues $\tilde{M}_{1}$ and $\tilde{M}_{2}$ are given
by 
\begin{equation}
\tilde{M}_{1,2}^{2}=\frac{1}{2}(A+\Sigma )\mp \frac{1}{2}[(\Delta _{0}\cos
2\theta -A)^{2}+\Delta _{0}^{2}\sin ^{2}2\theta ]^{\frac{1}{2}}+A_{n}
\label{139}
\end{equation}

where $\Sigma \equiv m_{2}^{2}+m_{1}^{2}$ and $\Delta _{0}\equiv
m_{2}^{2}-m_{1}^{2}$ . One can also obtain, by diagonalizing ${\cal M}%
_{-}^{2}$ the in-medium mixing angle $\theta _{M}$. The well-known result 
\begin{equation}
\sin ^{2}2\theta _{M}=\frac{\Delta _{0}^{2}\sin ^{2}2\theta }{(\Delta
_{0}\cos 2\theta -A)^{2}+\Delta _{0}^{2}\sin ^{2}2\theta }  \label{156}
\end{equation}

is then reproduced.

The procedure discussed in Appendix A can be extended in order to construct
the neutrino Wigner functions in the equilibrium state described by the
Hartree approximation. We will only give the final result. As before, it is
illustrative to consider first the case of one generation (for example,
electron neutrinos). One obtains 
\begin{equation}
F^{(\nu )}(p)=F^{-}(p)+F^{+}(p)+\bar{F}^{-}(p)+\bar{F}^{+}(p)  \label{Feq1}
\end{equation}

where 
\begin{eqnarray}
F^{-}(p) &=&(2\pi )^{-3}2E_{p}\delta (p^{2}-M_{-}^{2})f(p)\Sigma ^{-}(\vec{p}%
)  \nonumber \\
F^{+}(p) &=&(2\pi )^{-3}2E_{p}\delta (p^{2}-M_{+}^{2})f(p)\Sigma ^{+}(\vec{p}%
)  \nonumber \\
\bar{F}^{-}(p) &=&(2\pi )^{-3}2E_{p}\delta (p^{2}-\bar{M}_{-}^{2})\bar{f}(p)%
\bar{\Sigma}^{-}(\vec{p})  \nonumber \\
\bar{F}^{+}(p) &=&(2\pi )^{-3}2E_{p}\delta (p^{2}-\bar{M}_{+}^{2})\bar{f}(p)%
\bar{\Sigma}^{+}(\vec{p})  \label{Feq4}
\end{eqnarray}

and the effective masses are defined in Eqs. (\ref{mef1})-(\ref{mef4}). The
chiral left-handed component $F_{L}^{(\nu )}(p)$ , as defined in Eq. (\ref
{projection}) can be approximated by : 
\begin{equation}
F_{L}^{(\nu )}(p)\simeq F_{L}^{-}(p)+\bar{F}_{L}^{+}(p)=(2\pi )^{-3}2E_{p}%
\left[ \delta (p^{2}-M_{-}^{2})f(p)\Sigma _{L}^{-}(\vec{p})+\delta (p^{2}-%
\bar{M}_{+}^{2})\bar{f}(p)\bar{\Sigma}_{L}^{+}(\vec{p})\right]
\end{equation}
We have introduced the following definitions : 
\begin{eqnarray}
\Sigma _{L}^{-}(\vec{p}) &\equiv &P_{L}\Sigma ^{-}(\vec{p})P_{R}\simeq
-\left( 
\begin{array}{cc}
0 & 0 \\ 
1 & 0
\end{array}
\right) \rho (\vec{p}) \\
\bar{\Sigma}_{L}^{+}(\vec{p}) &\equiv &P_{L}\bar{\Sigma}^{+}(\vec{p}%
)P_{R}\simeq \left( 
\begin{array}{cc}
0 & 0 \\ 
1 & 0
\end{array}
\right) \mu (\vec{p})  \nonumber
\end{eqnarray}
\begin{eqnarray*}
\Sigma _{L}^{+}(\vec{p}) &\equiv &P_{L}\Sigma ^{+}(\vec{p})P_{R}\simeq 0 \\
\bar{\Sigma}_{L}^{-}(\vec{p}) &\equiv &P_{L}\bar{\Sigma}^{-}(\vec{p}%
)P_{R}\simeq 0
\end{eqnarray*}
The latter approximations hold for ultra-relativistic neutrinos. Thus, under
this approximation the field contains only two degrees of freedom :
neutrinos with negative helicity and antineutrinos with positive helicity.

We now go to the case of two neutrino generations. As in the free case, the
Wigner function will be diagonal if considered in the interaction
eigenstates basis : 
\begin{equation}
\widetilde{F}^{(\nu )ab}(p)=\left( 
\begin{array}{cc}
\tilde{F}^{\,(\nu )11}(p) & 0 \\ 
0 & \tilde{F}^{\,(\nu )22}(p)
\end{array}
\right)
\end{equation}
Moreover, each one of the diagonal components can be easily obtained by
taking into account the corresponding dispersion relations. One then arrives
to a set of equations similar to Eqs. (\ref{Feq1})-(\ref{Feq4}) for each one
of the two diagonal components of the Wigner function. As in the case of one
generation, we concentrate on the chiral left-handed component, which is
finally approximated by 
\begin{equation}
\widetilde{F_{L}}^{(\nu )ab}(p)\simeq \widetilde{F_{L}}^{-\,ab}(p)+%
\widetilde{\bar{F}_{L}}^{+ab}(p)=(2\pi )^{-3}2E_{p}\delta _{ab}\left[ \delta
(p^{2}-\tilde{M}_{a}^{2})f_{a}(p)\Sigma _{L}^{-}(\vec{p})+\delta (p^{2}-%
\widetilde{\bar{M}}_{a}^{2})\bar{f}_{a}(p)\bar{\Sigma}_{L}^{+}(\vec{p})%
\right] \qquad
\end{equation}
with $a,b=1,2$\ \ . The functions $f_{a}(p)$ and $\bar{f}_{a}(p)$ are the
same functions defined in Eqs. (\ref{f2gplus}) and (\ref{f2gminus}), but
with the replacement $m_{a}\rightarrow \tilde{M}_{a}$. The two terms in the
latter equation obey the relations 
\begin{eqnarray}
(\gamma p-\widetilde{M}^{ab})\widetilde{F}_{L}^{-\,ab}(p) &=&0  \nonumber \\
(\gamma p-\widetilde{\bar{M}}^{ab})\widetilde{\bar{F}}_{L}^{+\,ab}(p) &=&0
\end{eqnarray}
where $\widetilde{M}^{ab}$and $\widetilde{\bar{M}}^{ab}$are the (diagonal)
effective masses matrices for neutrinos and antineutrinos. In order to
construct the Wigner Function in the flavor space, we perform an unitary
transformation $U_{M}$ that is defined by the mixing angle in mater $\theta
_{M}$ . In this way we have, for example : 
\begin{eqnarray}
F_{L}^{-ee}(p) &=&\cos ^{2}(\theta _{M})\widetilde{F}_{L}^{-11}(p)+\sin
^{2}(\theta _{M})\widetilde{F}_{L}^{-22}(p)  \nonumber \\
F_{L}^{-e\mu }(p) &=&F_{L}^{-\mu e}(p)=\sin (\theta _{M})\cos (\theta _{M}) 
\left[ \widetilde{F}_{L}^{-11}(p)-\widetilde{F}_{L}^{-22}(p)\right] 
\nonumber \\
F_{L}^{-\mu \mu }(p) &=&\sin ^{2}(\theta _{M})\widetilde{F_{L}}%
^{-11}(p)+\cos ^{2}(\theta _{M})\widetilde{F_{L}}^{-22}(p)
\end{eqnarray}

\end{document}